\documentclass[aps,prb,twocolumn,superscriptaddress,draft]{revtex4}
\input{epsf}
\usepackage{amsmath}
\usepackage{amssymb}
\input{epsf}

\newcommand{\rem}[1]{}

\begin{document}

\title{Gaussian wave packets in phase space: The Fermi $g_F$ function}
\author{Giuliano Benenti}
\email{giuliano.benenti@uninsubria.it}
\affiliation{CNISM, CNR-INFM \& Center for Nonlinear and Complex Systems,
Universit\`a degli Studi dell'Insubria, Via Valleggio 11, I-22100 Como, Italy}
\affiliation{Istituto Nazionale di Fisica Nucleare, Sezione di Milano,
via Celoria 16, I-20133 Milano, Italy}
\author{Giuliano Strini}
\email{giuliano.strini@mi.infn.it}
\affiliation{Dipartimento di Fisica, Universit\`a degli Studi di Milano,
Via Celoria 16, I-20133 Milano, Italy}
\date{October 23, 2008}

\begin{abstract}
Any pure quantum state can be equivalently represented by means
of its wave function $\psi(q)$ or of the Fermi function $g_F(q,p)$,
with $q$ and $p$ coordinates and conjugate momenta of the system 
under investigation.
We show that a Gaussian wave packet can be conveniently visualized
in phase space by means of the curve $g_F(q,p)=0$. The evolution
in time of the $g_F=0$ curve is then computed for a Gaussian 
packet evolving freely or under a constant or a harmonic force.
As a result, the spreading or shrinking of the packet is 
easily interpreted in phase space.
Finally, we discuss a gedanken prism microscope experiment for measuring 
the position-momentum correlation. This gedanken experiment, 
together with the well-known Heisenberg microscope and 
von Neumann velocimeter, is sufficient to fully determine the 
state of a Gaussian packet.
\end{abstract}

\maketitle
 
\section{Introduction}
\label{sec:intro}

The free propagation of a Gaussian wave packet is the simplest 
application of the time-dependent Schr\"odinger equation. 
In their first quantum mechanics course, students soon learn
that, for a free particle, a minimum-uncertainty Gaussian packet 
spreads in time. They also learn that, for a harmonic oscillator,
such minimum packet does not spread. 
Since, according to the Ehrenfest theorem, a quantum
wave packet follows a beam of classical orbits as long as the packet 
remains narrow, one can conclude that the harmonic motion of a 
minimum packet can be described by means of classical mechanics.

On the other hand, it should be stressed that: 

(i) If we call $q$ and $p$ the coordinate and momentum of the system 
under investigation, then also a beam of classical orbits 
can spread along $q$. That is, the spreading 
is not necessarily a quantum phenomenon.

(ii) Gaussian wave packets can also contract along $q$, namely
the position uncertainty can narrow along $q$. Given an
arbitrarily long time $\tau$, it is 
possible to build packets that shrink up to that 
time.~\cite{klein}

Heisenberg uncertainty principle forbids the notion of
trajectories in quantum mechanics.
However, a comparison between classical and quantum dynamics can be made
by studying the evolution in time of the classical and quantum 
phase space distributions.~\cite{lee,heller,qcbook2,brumer,zurek,davidovich,habib} 
The classical phase space distribution $\rho_C(q,p,t)$ determines 
the probability $\rho_C(q,p,t)\,dq\,dp$ that a particle is 
found at time $t$ in the infinitesimal phase space volume 
$dq\,dp$ centered at $(q,p)$. 
The evolution in time of $\rho$ is governed by the 
Liouville's equation.~\cite{goldstein}
A convenient phase space formulation of quantum mechanics is
obtained using the Wigner distribution function 
$\rho_W(q,p,t)$.~\cite{wigner,wignerreview}
Such distribution evolves according to a 
quantum Liouville equation.
However, the Wigner function in general 
is not positive definite and 
therefore requires the introduction of delicate 
concepts such as that of quasiprobability. 

In this paper, we follow a different phase space approach,
based on an old paper by Fermi.~\cite{fermi} The relevant
quantity here is the Fermi $g_F$ function, or more 
precisely the $g_F(q,p,t)=0$ curve. We will
show that the $g_F$ function provides a nice intuitive 
phase space picture of Gaussian wave packets and of their 
evolution in time. In particular, the spreading or shrinking 
of a Gaussian packet evolving freely or under a constant
or a harmonic force are easily interpreted in the
Fermi-function picture. Finally,
we will discuss gedanken experiments determining 
the $g_F=0$ Fermi curve for Gaussian packets. In particular, 
we will focus on a prism microscope capable of measuring 
the position-momentum correlation. It turns out that 
the prism microscope, together with the well-known 
Heisenberg microscope and von Neumann's velocimeter, 
fully determines the $g_F=0$ curve for Gaussian packets.

\section{The Fermi $g_F$ function}

For the sake of simplicity, we consider the case 
of a single particle moving along a straight line.  
In classical mechanics, it is possible to determine the state of such system 
by measuring at some time $t$ the values of the position $q$ and momentum
$p$ or, equivalently, by measuring two independent functions 
$f(q,p)$ and $g(q,p)$, from which the values of $q$ and $p$ can be obtained.

In quantum mechanics, due to Heisenberg uncertainty principle,
a simultaneous exact measurement of $q$ and $p$ is not possible.
An exact measurement is in principle possible only for one physical
quantity: $q$ or $p$ or any function $g(q,p)$. Of course different
choices of $g(q,p)$ lead to different determinations of the system.
For instance, an exact measurement of $q$ implies a complete 
indetermination of $p$, and vice versa. 

As pointed out by Fermi,~\cite{fermi} the state of a quantum system at 
a given time $t$ may be defined in two completely equivalent ways: 
(i) by its wave function $\psi(q,t)$ or 
(ii) by measuring a physical quantity $g_F(q,p,t)$. 
Indeed, given the measurement outcome $g_F(q,p,t)=\bar{g}$, then 
$\psi(q,t)$ is obtained as solution of the eigenvalue equation 
$g_F(q,p,t)\psi(q,t)=\bar{g}\psi(q,t)$, where 
$p=-i\hbar{\partial_q}$.
On the other hand, given the wave function $\psi(q,t)$ 
it is always possible to find an operator $g_F(q,p,t)$ such that
\begin{equation}
g_F(q,p,t) \psi(q,t)=0.
\label{goperator}
\end{equation}
Using the polar decomposition 
\begin{equation}
\psi(q,t)=\rho(q,t)e^{i\theta(q,t)},
\end{equation}
where $\rho$ and $\theta$ are real ($\rho\ge 0$), it is easy to
check that identity (\ref{goperator}) is indeed fulfilled by taking 
\begin{equation}
{g}_F\left(q,-i\hbar\partial_q,t\right)=
\left[-i\hbar\partial_q -
\hbar\partial_q \theta(q,t)\right]^2+
\hbar^2\frac{\partial_{qq}^2 \rho(q,t)}{\rho(q,t)}.
\label{gfoperator}
\end{equation}

Equation (\ref{goperator}) implies that the corresponding 
physical quantity $g_F(q,p,t)$ takes the value $\bar{g}=0$. 
Therefore, the equation
\begin{equation}
g_F(q,p,t)=
\left[p-\hbar\partial_q \theta(q,t)\right]^2+
\hbar^2\frac{\partial_{qq}^2 \rho(q,t)}{\rho(q,t)}=0
\label{gfimplicit}
\end{equation}
defines a curve in the two-dimensional phase space,
parametrically dependent on time $t$.
In other words, as expected from Heisenberg uncertainty principle, 
we cannot identify a quantum particle by means of a point $(q,p)$
but we need a curve $g_F(q,p,t)=0$. We call $g_F$ the Fermi 
function and we are interested in the phase space curve $g_F=0$.
It is also possible to write equation (\ref{gfimplicit}) in the form
\begin{equation}
p_\pm=\hbar\partial_q \theta(q,t)\pm
\sqrt{-
\hbar^2\frac{\partial_{qq}^2 \rho(q,t)}{\rho(q,t)}}.
\label{ppmFermi}
\end{equation}
This equation locates two points $(q,p_+)$ and $(q,p_-)$ in the 
phase space for any $q$ such that $\partial_{qq}^2 \rho(q,t)<0$
and $\rho(q,t)\ne 0$.~\cite{footnote}

\section{The Fermi $g_F$ function for Gaussian wave packets}

In this section, we discuss the phase space curve $g_F(q,p,t)=0$
corresponding to the wave function $\psi(q,t)$ solution to the
one-dimensional Schr\"odinger equation  
\begin{equation}
i \hbar  \partial_t  \psi(q,t)=
\left[-\frac{\hbar^2}{2m}  {\partial_{qq}^2} + V(q)\right]
\psi(q,t),
\label{schrodinger1d}
\end{equation}
for various time-independent potentials $V(q)$ such that
an initially Gaussian wave packet
$\psi(q,0)$ remains Gaussian at all times.

\subsection{Free particle}

\begin{figure} 
\centerline{\epsfxsize=8.5cm\epsffile{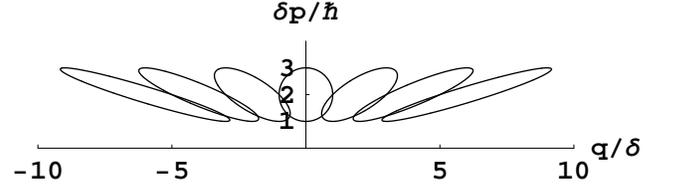}}
\caption{Phase space curves $g_F=0$ at different times 
(from left to right, 
$\tau=\frac{\hbar t}{m \delta^2}=-3,-2,-1,0,1,2,3$) 
for a free particle, whose state 
at $t=0$ is a minimum uncertainty Gaussian wave packet centered at 
$\left(\frac{q_0}{\delta}=0,\frac{\delta p_0}{\hbar}=2\right)$.}
\label{fig:freeparticle}
\end{figure}

We have $V(q)=0$ and consider as initial condition the Gaussian
minimum uncertainty wave packet
\begin{equation}
\psi(q,0)=
\frac{1}{\sqrt{\sqrt{\pi}\delta}}
\exp\left\{-\frac{(q-q_{0})^{2}}{2\delta^{2}}+
\frac{i}{\hbar}[ p_{0}(q-q_{0})]\right\}.
\label{gaussinitial}
\end{equation}
The solution to the Schr\"odinger equation (\ref{schrodinger1d})
with this initial condition 
reads
\begin{eqnarray}
\begin{array}{c}
{\displaystyle
\psi(q,t) =
  \frac{1}{\sqrt{\sqrt\pi\left(1+ i \frac{\hbar t}{m\delta^2}\right)\delta}}
  \exp\left\{
  -\frac{\left( q - q_0 - \frac{p_0 t}{m} \right)^2}
        {2\delta^2\left(1+i\frac{\hbar t}{m\delta^2}\right)}
\right.
}
\\
\\
{\displaystyle
\left.
    + \frac{i}\hbar    \left[
      p_0 (q - q_0) - \frac{p_0^2 t}{2m} 
    \right]
  \right\}.
}
\end{array}
\end{eqnarray}
We now apply operator (\ref{gfoperator}) to this wave function,
thus obtaining
\begin{equation}
f(t)g_F(q,p,t)=
\frac{\tilde {q}^{2}}{\delta^{2}}
+\frac{\delta^{2}}{\hbar^{2}}
\left(1+\frac{\hbar^{2}t^{2}}{m^{2}\delta^{4}}\right)\tilde{p}^2
-\frac{2t}{m\delta^2}
\tilde{q}\tilde {p} -1,
\label{gffreeparticle}
\end{equation}
where for later convenience we have multiplied the Fermi function by
$f(t)=\frac{\delta^2}{\hbar^2}\left(1+\frac{\hbar^2 t^2}{m^2\delta^4}\right)$
and we have defined
\begin{equation}
\tilde {q} = q - q_{0}- \frac{p_{0}t}{m},
\quad
\tilde {p}=p-p_{0}.
\end{equation}
The $g_F(q,p,t)=0$ curve is the ellipse
\begin{equation}
a \tilde{q}^2+b\tilde{p}^2+2c\tilde{q}\tilde{p}=1,
\label{ellipse}
\end{equation} 
where 
\begin{equation}
a=\frac{1}{\delta^2},\quad
b=\frac{\delta^2}{\hbar^2}
\left(1+\frac{\hbar^2 t^2}{m^2\delta^4}\right),\quad
c=-\frac{t}{m\delta^2}. 
\end{equation}
As shown in Fig.~\ref{fig:freeparticle}, the shape of 
the ellipse changes over time. However, its area 
\begin{equation}
A=\frac{\pi}{\sqrt{ab-c^2}}=\pi\hbar
\end{equation}
remains constant.

\subsection{Uniformly accelerating particle}

\begin{figure} 
\centerline{\epsfxsize=8.5cm\epsffile{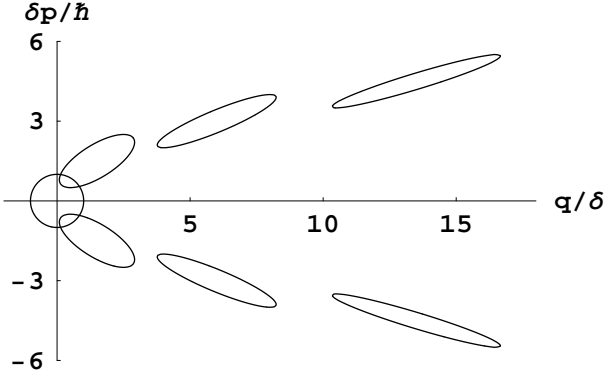}}
\caption{Same as in Fig.~\ref{fig:freeparticle}, but for
a uniformly accelerating particle with  $\frac{m \delta^3F_0}{\hbar^2}=1.5$. 
The $t=0$ minimum uncertainty Gaussian wave packet is centered at
$\left(\frac{q_0}{\delta}=0,\frac{\delta p_0}{\hbar}=0\right)$.}
\label{fig:acceleratingparticle}
\end{figure}

We have $V(q)=-F_0 q$. In this case, the solution to the 
Schr\"odinger equation (\ref{schrodinger1d}) with initial 
condition the Gaussian minimum uncertainty wave packet
(\ref{gaussinitial}) reads
\begin{eqnarray}
\begin{array}{c}
{\displaystyle
\psi(q,t)= }\\
{\displaystyle
  \frac{1}{\sqrt{\sqrt\pi\left(1+ i \frac{\hbar t}{m\delta^2}\right)\delta}}
  \exp\left\{
  -\frac{\left( q - q_0 - \frac{p_0 t}{m} -\frac{F_0 t^2}{2m}\right)^2}
        {2\delta^2\left(1+i\frac{\hbar t}{m\delta^2}\right)}
\right. 
}
\\
\\
{\displaystyle
\left.
    + \frac{i}\hbar    \left[
      p_0 (q - q_0) - \frac{p_0^2}{2m}  t + \phi
    \right]
  \right\},
}
\end{array}
\end{eqnarray}
where
\begin{equation}
\phi=F_0 qt-\frac{F_0 p_0 t^2}{2 m} -\frac{F_0^2 t^3}{6m}.
\end{equation}
The Fermi $g_F$ function is again given by equation 
(\ref{gffreeparticle}), but with the center of the ellipse 
accelerating uniformly, namely we have
\begin{equation}
\tilde {q} = q - q_{0}- \frac{p_{0}t}{m}-\frac{F_0 t^2}{2m},
\quad
\tilde {p}=p-p_{0}-F_0 t.
\end{equation}
The deformation with time of the ellipse $g_F=0$
is independent of the force $F_0$, which simply displaces the 
center $(\tilde{q},\tilde{p})$ of the ellipse along the parabola
\begin{equation}
q=q_0+\frac{p_0}{mF_0}(p-p_0)+\frac{(p-p_0)^2}{2mF_0}.
\end{equation} 
The curves $g_F=0$ at different times are shown in 
Fig.~\ref{fig:acceleratingparticle}. 

\subsection{Harmonic oscillator}

We have $V(q)=\frac{1}{2}m\omega_0^2q^2$. A Gaussian solution 
to the Schr\"odinger equation (\ref{schrodinger1d}) is given by 
\begin{equation}
\psi(q,t)=N(t)\exp\left\{-\frac{1}{2}A(t)[q-Q(t)]^2
+i\chi(q,t)\right\},
\label{gaussharmonic}
\end{equation}
where 
\begin{equation}
N(t)=\sqrt{\frac{\alpha}{\sqrt{\pi}}}\frac{1}{\sqrt{C(t)+i\sqrt{B}S(t)}},
\label{hh1}
\end{equation}
\begin{equation}
C(t)=\cos(\omega_0 t +\phi),
\;
S(t)=\sin(\omega_0 t +\phi),
\;
B=\frac{\hbar^2 \alpha^4}{m^2\omega_0^2},
\label{hh2}
\end{equation}
\begin{equation}
A(t)=A_r(t)+iA_i(t),
\label{hh3}
\end{equation}
\begin{equation}
A_r(t)=\frac{m\omega_0\sqrt{B}}{\hbar[C^2(t)+BS^2(t)]},
\end{equation}
\begin{equation}
A_i(t)=\frac{m\omega_0(1-B)C(t)S(t)}{\hbar[C^2(t)+BS^2(t)]},
\label{hh4}
\end{equation}
\begin{equation}
Q(t)=Q_0 C(t),
\end{equation}
\begin{equation}
\chi(q,t)=-\frac{\omega_0 m}{\hbar} Q_0 q S(t)+
\frac{m \omega_0}{2 \hbar} Q_0^2 C(t) S(t).
\label{hh5}
\end{equation}
Different values of the parameters $\alpha$, $Q_0$ and $\phi$
correspond to different solutions to the Schr\"odinger equation.
Note that, for the sake of simplicity, we have not considered in 
(\ref{gaussharmonic}) the most general Gaussian wave packet for
the harmonic oscillator.~\cite{dodonov} 

\begin{figure}
\centerline{\epsfxsize=8.5cm\epsffile{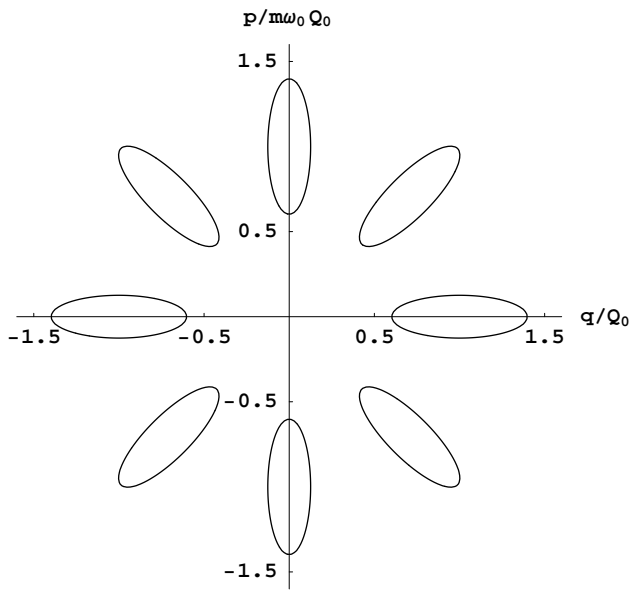}}
\caption{Same as in Fig.~\ref{fig:freeparticle}, but for
the squeezed state (\ref{gaussharmonic}) of a harmonic oscillator,
with $\frac{m\omega_0 Q_0^2}{\hbar}=20$, $B=0.1$
and $\omega_0 t+\phi$ ranging from $0$ to $7\pi/4$, in steps of $\pi/4$.}
\label{fig:harmonic}
\end{figure}

Due to the time dependence of $A(t)$, it is not immediate to 
grasp the evolution of the Gaussian packet (\ref{gaussharmonic}).
A clear illustration is provided by the Fermi function.
The $g_F(q,p,t)=0$ curve is derived from equation (\ref{gfimplicit}) 
and again is a (parametrically dependent on $t$) ellipse in phase space,
whose equation can be written in the form (\ref{ellipse}), with 
\begin{equation}
a=\frac{A_r^2+A_i^2}{A_r},
\quad
b=\frac{1}{\hbar^2 A_r},
\quad
c=\frac{A_i}{\hbar A_r},
\label{ellipsesqueezed}
\end{equation}
\begin{equation}
\tilde{q}=q-Q(t),\quad
\tilde{p}=p-m\dot{Q}(t).
\end{equation}
Note that the area of the ellipse remains constant: $A=\pi\hbar$.
Owing to the term proportional to $\tilde{q}\tilde{p}$, the ellipse
in general does not have its axes parallel
to the coordinate axes $q,p$. This term disappears when 
$A_i=0$, that is, for $B=1$ ($\alpha^2=\frac{m\omega_0}{\hbar}$).
This special case corresponds to coherent states, for which 
the $g_F=0$ curve becomes, in the coordinate plane 
$\left(q,\frac{p}{m\omega_0}\right)$, a rigidly moving circle:
\begin{equation}
[q-Q(t)]^2+\frac{[p-m\dot{Q}(t)]^2}{m^2\omega_0^2}=
\frac{\hbar}{m\omega_0}.
\end{equation}
A more generic example (squeezed state) is shown in Fig.~\ref{fig:harmonic}.

\section{Measuring the position-momentum correlation: 
A gedanken experiment}

It is easy to check that,
for the examples discussed above, the evolution of the phase space
region enclosed by the $g_F(q,p,t)=0$ curve is the same as for the 
corresponding classical phase space distribution $\rho_C(q,p,t)$, 
uniform inside such region and zero-valued outside 
and evolving according to the 
Liouville equation. Due to this equivalence, the conservation of the
area of this region is assured by the Liouville theorem.~\cite{goldstein}
However, we should stress that the spreading or the shrinking of
a wave packet are quantum effects, in the sense that the wave
packet refers to a single particle, while the distribution 
function $\rho_C(q,p,t)$ describes a classical ensemble of orbits. 
No physical meaning should be attached to each point inside the
$g_F=0$ curve. Due to Heisenberg principle it is not
possible to assign with arbitrarily small uncertainties both 
position and momentum.

For Gaussian states, the area enclosed by the
Fermi $g_F$ function has a simple 
interpretation in terms of 
Heisenberg uncertainty principle. To illustrate this point, let us
compute the variances
\begin{equation}
(\Delta q)^2=\langle (q-\langle q \rangle)^2\rangle, 
\quad
(\Delta p)^2=\langle (p-\langle p \rangle)^2\rangle
\end{equation}
and the position-momentum  correlation
\begin{equation}
K=\frac{1}{2}\langle (q-\langle q \rangle)(p -\langle p \rangle)
+(p-\langle p \rangle)(q -\langle q \rangle)\rangle
\end{equation}
for the examples discussed in the previous section.
In the case of a free or uniformly accelerating particle we obtain
\begin{equation}
(\Delta q)^2=\frac{\delta^2}{2}\left(1+\frac{\hbar^2 t^2}{m^2\delta^4}\right),
\;
(\Delta p)^2=\frac{\hbar^2}{2\delta^2},
\;
K=\frac{\hbar^2 t}{2m\delta^2},
\end{equation}
while for the harmonic oscillator
\begin{equation}
\begin{array}{c}
{\displaystyle
(\Delta q)^2=\frac{1}{2A_r},
\;
(\Delta p)^2=\frac{\hbar^2(A_r^2+A_i^2)}{2A_r},
}
\\
{\displaystyle
K=-\frac{\hbar(1-B)CS}{2\sqrt{B}}.}
\end{array}
\end{equation}
In both cases, the generalized uncertainty 
relation~\cite{robertson,dodonov,sudarshan}
\begin{equation}
(\Delta q)^2 (\Delta p)^2 -K^2=\frac{\hbar^2}{4}
\label{generalizeduncertainty}
\end{equation}
is fulfilled. Moreover, the parameters 
determining the $g_F=0$ ellipse can be expressed 
as follows:
$a=\frac{2}{\hbar^2}(\Delta p)^2$,
$b=\frac{2}{\hbar^2}(\Delta q)^2$ and
$c= \frac{2}{\hbar^2}K$ (for the free or uniformly accelerating
particle) or $c= -\frac{2}{\hbar^2}K$ (for the harmonic oscillator).
This implies that in these examples the experimental determination of 
the $g_F=0$ ellipse is possible once the quantities
$\langle q \rangle$, $(\Delta q)^2$, $\langle p \rangle$,
$(\Delta p)^2$ and $K$ are measured. 

A simple consequence is that the spreading (in $q$)
of a Gaussian wave packet for a free or uniformly accelerating particle 
as well as the oscillations in time of the $q$ and $p$ variances for a 
harmonic oscillator can be simply visualized in terms of the evolution
in time of the $g_F=0$ curve or, equivalently, of
the phase space region enclosed by this curve.
We stress that here the conservation of the area of the 
$g_F=0$ ellipse is equivalent to the validity of the generalized 
uncertainty relation (\ref{generalizeduncertainty}) at all times.
The product $\Delta q \Delta p$ may grow or even reduce in time.
This can be clearly seen from Figs.~\ref{fig:freeparticle} and
\ref{fig:acceleratingparticle}: $\Delta q \Delta p$ grows if
we start from a minimum uncertainty state; on the other hand, if 
we start from a different state, $\Delta q \Delta p$ may decrease
until the minimum uncertainty state is reached and then starts
growing.

Notwithstanding the behaviour of $\Delta q\Delta p$, the uncertainty in 
the measurement of the state in 
phase space is constant, provided that we measure not
only position and momentum but also their correlation $K$.
For the complete determination of the $g_F=0$ ellipse we need
three thought experiments:

(i) The measurement of $\langle q \rangle$ and 
$(\Delta q)^2$. This can be done by means of Heisenberg
microscope.

(ii) The mesurement of $\langle p \rangle$ and 
$(\Delta p)^2$, possible using von Neumann velocimeter.

(iii) The measurement of the position-momentum correlator $K$. 

\begin{figure}
\centerline{\epsfxsize=8.5cm\epsffile{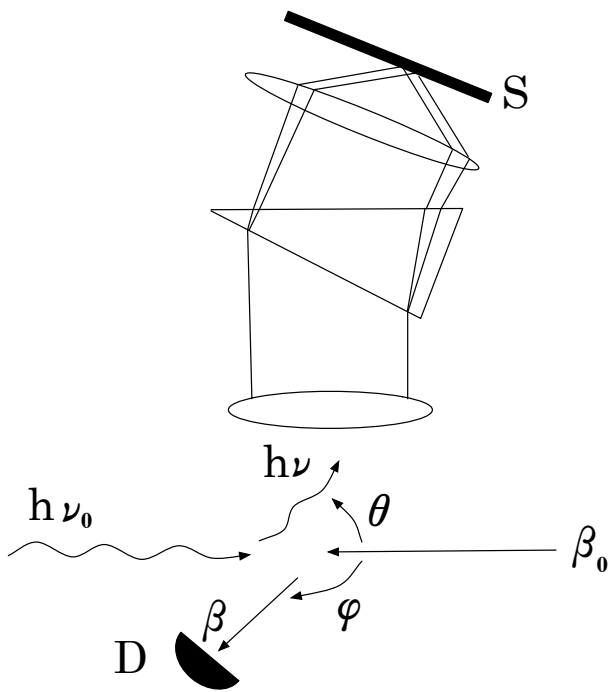}}
\caption{Schematic drawing of a thought microscope, measuring the 
position-momentum correlation of a particle. The scattered light
is collected on the screen $S$. Note that a coincidence system
with the detector $D$ of the scattered particle is necessary,
in order to fix the scattering angle.} 
\label{fig:microscope}
\end{figure}

The first two gedanken experiments are well known and their
description can be found, for instance, in Ref.~\cite{vonNeumann}.
In order to measure the correlator $K$, one can use  
Heisenberg microscope with the addition of a prism after 
the objective,~\cite{grassistrini} see Fig.~\ref{fig:microscope}.

The main idea of this prism microscope gedanken experiment
is suggested by the objective prism spectroscopy used 
for astronomical spectroscopy. In that case, a prism is placed in
front of the objective lens of a telescope. As a result, the light
coming from stars within the field of view of the telescope 
is dispersed by the prism depending on its frequency. Therefore,
the telescope can be used for simultaneously measuring the spectra
of many stars.~\cite{physicstoday}
In our case, the prism is located after the lens of the microscope
(see Fig.~\ref{fig:microscope}), so that, if the refractive 
index of the prism is a suitable function
of the wave length of the scattered photon, then such
photon strikes the screen $S$ in a position dependent of the
position-momentum correlation of the measured particle.

We now sketch the working of the prism microscope.
We first write the equations of the Compton effect for the
scattering between the photon (our probe) and the measured particle. 
Considering a scattering event on the plane of Fig.~\ref{fig:microscope},
the nonrelativistic energy-momentum conservation laws read as follows:
\begin{equation}
\left\{
\begin{array}{l}
\displaystyle
h \nu_0 +\frac{1}{2} m c^2 \beta_0^2 =
h \nu +\frac{1}{2} m c^2 \beta^2, 
\\
\displaystyle
h\nu_0-m \beta_0 c^2 =
h\nu \cos \theta +m \beta c^2 \cos \varphi,
\\
\displaystyle
h\nu \sin \theta - m \beta c^2 \sin \varphi=0,
\end{array}
\right.
\label{eq:compton}
\end{equation}
where $h$ is Planck constant, $c$ the speed of light in vacuum, 
$m$ the particle mass,
$\nu_0$ and $\nu$ ($\beta_0 c$ and $\beta c$)
denote the photon frequency (the particle velocity) 
before and after the collision, 
$\theta$ and $\varphi$ are the scattering angles.

There are five unknown parameters in the Compton scattering equations
(\ref{eq:compton}): $\nu,\beta_0,\beta,\theta,\varphi$. 
If we fix $\varphi$ and exploit (\ref{eq:compton}), then 
we can obtain a single equation $F(\nu,\beta_0)=0$ 
that relates the frequency $\nu$
of the scattered light to the velocity $c\beta_0$ of the 
particle. By linearizing such equation we obtain 
\begin{equation}
\nu\approx A + B (\beta_0-\langle \beta_0 \rangle), 
\label{eq:comptonlinear}
\end{equation}
with $A,B$ constants
and $c\langle \beta_0 \rangle$ mean value of the particle velocity.
It is clear that, in order to fix the angle $\varphi$, a coincidence 
system between the detector measuring the angle $\varphi$ of the scattered 
particle and the screen is needed.

If the refractive index of the prism is a suitable and known function 
of the frequency $\nu$ of the scattered light and assuming that
linearization (\ref{eq:comptonlinear}) holds, then we collect on a
screen a spot whose location depends on the position-momentum 
correlation of the measured particle.
If without the prism the photon hits the screen at 
the position $\xi=q$, then with the prism $\xi$ is displaced
by an amount depending on the frequency $\nu$. If the refractive index 
is a linear function of $\nu$, then we obtain $\xi=q+ C p$,
with $C$ constant. 
Considering a more generic suitable nonlinear dependence of the 
refractive index on $\nu$, we obtain 
\begin{equation}
\xi\approx q + C p + D (qp+pq),
\label{eq:refractiveindex}
\end{equation}
with $D$ constant.
Since, as we have said above, $\langle q \rangle$ and $\langle p \rangle$ 
can be measured by means of other gedanken experiments, then the prism 
microscope is in principle suitable for the measurement of the correlator
$\langle qp+pq \rangle$.

It is of course understood that experiments (i)-(iii) must be 
repeated many times, with identically prepared wave functions,
to measure the above quantities $\langle q \rangle$,
$(\Delta q)^2$, $\langle p\rangle$, $(\Delta p)^2$ and $K$
with sufficient statistical accuracy.

\section{Conclusions}

We have shown that the Fermi $g_F$ function provides a very simple
and nice phase space visualization of the dynamics 
of Gaussian wave packets.
In particular, the area enclosed by the $g_F=0$ curve is of the
order of the Planck constant and is a conserved quantity.
We have also discussed a gedanken experiment for measuring the 
position-momentum correlation. This experiment adds to the 
well-known Heisenberg microscope and von Neumman velocimeter,
thus allowing the full determination of the state of 
a Gaussian wave-packet.   

The extension of the results obtained in this 
paper to more complex, non-Gaussian states has to face both 
numerical and conceptual difficulties. 
It is indeed clear from Eq.~(\ref{ppmFermi}) that numerical
errors in determining the $g_F=0$ curve become relevant
in the asymptotic region of large $q$ where the wave function 
amplitude $\rho$ is small. While for a Gaussian packet the 
$g_F=0$ curve is sufficient to fully determine the state 
of the system, this is not the case for a generic and a priori
unknown wave packet.
The complete determination
of the generic state of a system requires  
consideration of the complex values of $p_{\pm}$, obtained
from Eq.~(\ref{ppmFermi}) when $\partial_{qq}^2 \rho >0$.
We then obtain
\begin{equation}
\hbar\partial_q\theta = \frac{p_++p_-}{2},
\quad
\hbar^2\frac{\partial_{qq}^2\rho}{\rho}=
-\left(\frac{p_+-p_-}{2}\right)^2,
\end{equation}
from which the $g_F$ operator (\ref{gfoperator}), and consequently
the wave function $\psi(q,t)$ are determined.

With regard to the comparison with well-know phase space 
distributions, notably the Wigner function 
$\rho_W(q,p,t)$~\cite{wigner,wignerreview}, we first note
that for Gaussian packets the $g_F=0$ curve coincide with the level 
curve of the points $(q,p)$ such that 
$\rho_W(q,p,t)=\frac{1}{e}\rho_W^{\rm max}$, 
with $\rho_W^{\rm max}$ maximum
value of $\rho_W$.
Different contour levels of $\rho_W$ correspond
to different ``equipotential curves'' $g_{F}={\rm constant}$. 
However, the $g_F$ function is conceptually very different from the Wigner
function. First of all, not only $g_F$ fulfills Eq.(\ref{goperator}) but 
also other operators such as $g_F^2$, $g_F^3$,... satisfy 
$g_F^2\psi=0$, $g_F^3\psi=0$,... . 
All information is enclosed in the zeroes of the Fermi function, 
from which one can reconstruct the wave function $\psi$.
Finally, for the $g_F$ we cannot conceive any interpretation in terms
of quasiprobabilities as for the Wigner function.

The great simplicity of the Gaussian cases is 
due to the fact that the phase space region enclosed by 
the $g_F(q,p,t)=0$ curve evolves in time as the corresponding 
classical phase space distribution $\rho_C(q,p,t)$, uniform inside 
such region and zero-valued outside and whose dynamics is governed 
by the Liouville equation. 
In studying the free evolution of the superposition 
of two Gaussian wave packets we have seen that, as expected, this 
quantum-classical correspondence is broken. 
Due to quantum interference the $g_F=0$ curve exhibits a 
rich, non-classical structure and, in particular, the 
area of the region enclosed by this curve does not remain constant 
over time. 
Therefore, quantum interference effects might be studied
in phase space by means of the $g_F$ Fermi function.

Finally, we point out that it is possible to derive equations 
governing the time evolution of the $g_F=0$ curve~\cite{posilicano}.
Such equations for generic, non-Gaussian wave packets differ
from the classical Hamiltonian flow by the addition of $\hbar$-dependent
terms. Therefore, one could use the Fermi function to study the 
quantum-classical correspondence. According
to the Ehrenfest theorem, the propagation of a quantum mechanical
wave packet is described for short times by classical equations
of motion. The time at which this correspondence breaks down
is called the Ehrenfest time~\cite{berman}. 
After that time, due to quantum interference 
an initially Gaussian wave packet is ``destroyed'', in that it 
splits into new small packets~\cite{qcbook2}. Such splitting 
may be detected also by means of the Fermi function. 

\acknowledgments
The authors wish to thank Andrea Posilicano for several 
interesting and stimulating discussions.


\begin{thebibliography}{}
\bibitem{footnote}
Equation (\ref{ppmFermi}) can be
written in the form
\begin{equation}
p_\pm = m v_M \pm \sqrt{2mV_Q},
\end{equation}
where
\begin{equation}
v_M\equiv \frac{\hbar}{m}\partial_q \theta,\;\;
V_Q\equiv -\frac{\hbar^2}{2m}\frac{\partial_{qq}^2}{\rho},
\end{equation}
are known as the Madelung's velocity~\cite{madelung} 
and the ``quantum-mechanical'' 
potential~\cite{bohm}, 
respectively.
\end{thebibliography}
\end{document}